\documentclass[preprint, amsmath]{revtex4}
\usepackage{amssymb}
\usepackage{indentfirst}
\usepackage[mathscr]{eucal}
\usepackage{graphicx}

\begin{document}

\title{
``Ping-pong'' electron transfer. \\
II. Multiple reflections of the Loschmidt echo \\
and the wave function trapping by an acceptor.
      }

\author{
V.N. Likhachev, T.Yu. Astakhova, G.A. Vinogradov${}^*$
       }

\affiliation{
Emanuel Institute of Biochemical Physics, \\
 Russian Academy od Sciences, Moscow, Russian Federation
            }

\begin{abstract}

This paper continues the preceding paper on the problem of quantum
dynamics on the lattice. Firstly we consider the multiple
reflections of the wave function (Loschmidt echo). The phenomenon
of wave function concentration on the impurity site after
reflections is found. The solution representing the total
amplitude $a(t)$ is obtained as the series in terms of partial
amplitudes $a_k(t)$. The contribution of $k$th partial amplitude
becomes dominant only after $k$th reflection from the lattice end.
An excellent agreement between analytical and accurate numerical
results is obtained. Next problem, -- wave packet trapping by
defects, is solved by numerical simulation. Analytical expressions
are derived in few cases allowing to estimate the quantum
efficiency of charge transfer. Obtained results can qualitatively
explain recent experiments on the highly efficient charge
transport in olygonucleotides and polypeptides.

\vspace{2 cm}



\noindent ${}^*$The corresponding author:
\texttt{gvin@deom.chph.ras.ru}.

\end{abstract}

\maketitle


\section{Setting up a problem}

For the completeness we repeat few details of the setting up the
problem from the preceding paper, but in a somewhat different
notations.

We consider quantum system consisting of $(N+1)$ sites. Most left
site is an impurity site; other $N$ sites are reservoir. This
system is defined by the $(N+1) \times (N+1)$ matrix hamiltonian
\begin{equation}
  \label{Ham1}
  H =
\begin{pmatrix}
   E                        & \overrightarrow{{\bf v}}  \\
   \overrightarrow{{\bf v}} & \widehat {\bf B} \\
\end{pmatrix},
\end{equation}
where $E$ is the on-site energy of the impurity site; $\widehat
{\bf B}$ is the $N \times N$ tridiagonal matrix of reservoir with
matrix elements  $B_{i,j} = \delta_{i,i+1} + \delta_{i,i-1}$;
$\overrightarrow{{\bf v}}$ is the interaction of the impurity site
with the reservoir and is chosen in the form  $v_i = C
\delta_{i,1}$. The wave function is  $\overrightarrow{\Psi}(t) =
\left\{ a(t), \overrightarrow{\bf b}(t) \right\} = a(t), b_1(t),
b_2(t), \ldots, b_N(t)$. Initial condition: $a(t=0) = 1, \,\,
\overrightarrow{{\bf b}}(t=0) = 0$. The aim of this paper is to
find the amplitude $a(t)$ on the impurity site as the result of
multiple pave packet reflections (Loschmidt echo).

The dimensionless $(\hbar = 1)$ Schr\"odinger equation is
\begin{equation}
  \label{1.1}
  \left\{
  \begin{split}
  i \dfrac{ {\rm d} a(t)}{{\rm d} t} =
     & E a(t) + (\overrightarrow{\bf v} \cdot \overrightarrow{\bf
     b});
    \qquad \qquad a(t=0) = 1 \\
   i \dfrac{ {\rm d} \overrightarrow{\bf b}(t)}{{\rm d} t} =
     & \left( \widehat{\bf B} \cdot \overrightarrow{\bf b}(t) \right) +
   \overrightarrow{\bf v} \, a(t); \qquad  \overrightarrow{\bf b}(t=0) = 0.
  \end{split}
  \right.
\end{equation}

If vector $\overrightarrow{\bf b}$ is expanded in terms of the
eigenfunctions $b_k(i)$ of matrix $\widehat{\bf B}$, then the
following integro-differential equation can be obtained:
\begin{equation}
  \label{1.2}
   \dot a = - i E a - \int\limits_0^t
   B^N (t - \tau) \, a(\tau) \, {\rm d} \tau, \qquad a(t=0) = 1,
\end{equation}
with the kernel
\begin{equation}
  \label{1.3}
   B^N (t) = \sum\limits_{k=1}^{N}
   \exp(- i \varepsilon_k t)
   \left( \overrightarrow{\bf v} \cdot
   \overrightarrow{\bf b_k} \right)^2
\end{equation}
and $\varepsilon_k$ is $k$th eigenvalue.

The Schr\"odinger equations \eqref{1.1} for this model are:
\begin{equation}
  \label{1.4}
  \left\{
  \begin{split}
  i \dot a =   & \, E a + C b_1 \\
  i \dot b_1 = & \, b_2 + C a \\
  i \dot b_2 = & \, b_1 + b_3 \\
 \cdots  \,\,\,\,      & \, \,\,\, \cdots \cdots \cdots   \\
  i \dot b_N = & \, b_{N-1} \\
  \end{split}
  \right.
\end{equation}
with initial conditions $a(0) = 1, \,\, b_{i=1,2, \ldots,N}(0) =
0$. Eigenfunctions $\varepsilon(k)$ and eigenvalues $b_i(k)$ of
the matrix $\widehat{\bf B}$ are well known:
\begin{equation}
  \label{1.5}
  \varepsilon(k) = 2 \, \cos \left( \dfrac{\pi k}{N+1} \right);
  \quad
  b_k(i) = \sqrt{\dfrac{2}{N+1}} \sin \left( \dfrac{\pi k}{N+1}i
  \right) \qquad k = 1,2, \ldots, N.
\end{equation}

The kernel $B^N(t)$ of Eq.~\eqref{1.2}, in accordance with
\eqref{1.5}, is defined by the following sum:
\begin{equation}
  \label{1.6}
  B^N(t) = \dfrac{2 \, C^2}{N+1}
  \sum\limits_{k=1}^N \sin^2 \left( \dfrac{\pi k}{N+1} \right) \,
  \exp\left[ - 2 i\cos \left( \dfrac{\pi k}{N+1} \right)  t \right].
\end{equation}
%


\section{Expansion of $a(t)$ in terms of partial amplitudes}

In the preceding paper we have shown that the well formed impulse
with sharp forward front is generated at large times. The velocity
of the impulse front $v =2$ is the maximal group velocity. If a
lattice consists of $N$ sites, then the time of the first
returning to the impurity site is $t \approx N$. And the cycle of
$k$ returnings takes $ t \approx kN$.

In this paper the detailed analysis of amplitude $a(t)$ in the
time range $t \lesssim N$ was done. Now our goal is to find
amplitude $a(t)$ on the impurity site at all times. We solve
equation \eqref{1.2} with the kernel given by \eqref{1.6}.

Bearing this in mind, the solution of \eqref{1.2} is searched as
the expansion in terms of partial amplitudes $a_k(t)$. Each
amplitude $a_k(t)$ is negligible in the time range $t \lesssim kN$
and its contribution to the total amplitude $a(t)$ becomes
essential starting only from times $t \sim kN$.

In order to make sure in this possibility, the kernel $B^N(t)$ of
Eq.~ \eqref{1.6} should be transformed according to the Poisson
formula:
\begin{equation}
  \label{2.1}
  \begin{split}
  B^N(t) =  &\,  C^2 \sum\limits_{m = -\infty}^{m = + \infty} B_m(t),
    \quad {\rm where} \\
  B_m(t) =  & \, \dfrac{2}{\pi} \int\limits_0^{\pi} {\rm d} x
  \sin^2(x) \exp\left\{-2i [m(N+1) + t \cos(x)]  \right\}.
  \end{split}
\end{equation}

Note that $B_m(t)$ is the real function and there exists the
explicit expression for $B_0(t)$ through the Bessel functions:
\begin{equation}
  \label{2.2}
  B_0(t) = J_0(2t) + J_2(2t).
\end{equation}

The original equation \eqref{1.2} with regard to transformation
\eqref{2.1} is rewritten in the form:
\begin{equation}
  \label{2.3}
  \dot a(t) = - i E a(t) - C^2 \sum\limits_{m = -\infty}^{\infty}
  \int\limits_0^t B_m(t - t') \, a(t') \, {\rm d} t'.
\end{equation}

Note an important and useful property of functions $B_m$: every
function $B_m(t)$ is small in the time range $0 < t < mN$, such
that
\begin{equation}
  \label{2.4}
  B_m(t) \approx \dfrac{\sin(2t)}{\pi [m(N+1)]^3}, \quad
  m(N+1) - t \gg 1.
\end{equation}

With a reasonable degree of accuracy one can say that the function
$B_m(t) \approx 0$ when $t < mN$. Consequently, when $t < mN$,
then only terms $B_k$ with indices $k < m$ are essential in the
governing equation \eqref{2.3}.

Now we represent amplitude $a(t)$ as the sum of partial amplitudes
\begin{equation}
  \label{2.5}
  a(t) = a_0(t) + a_1(t) + a_2(t) + \ldots \,.
\end{equation}
Recall that our assumption concerning amplitudes $a_k(t)$ lies in
the fact that amplitude $a_k(t)$ is small at $t < kN$. Their
contribution to the total amplitude $a(t)$ becomes essential
starting from time $t \sim kN$. To some extent $a_k(t)$ behaves
like $B_k(t)$. Deriving an equation for $a_k(t)$, we assume that
its right-hand side contains only terms $B_m(t)$ with $m\leq k$.

Prior to write down an equation for $a_k(t)$, we make one comment.
Terms $B_m(t)$ are small (even for not too long lattices). This
smallness is ensured by the fact that $B_{-m}(t)= B_m(-t)$. And
$B_m(t) \sim (mN)^{-3}$ in the considered time range ($t>0$).
Nevertheless these terms are taken into account as their
accounting allows to derive explicit analytical formulaes.

Lets introduce following functions to account the contributions
from $B_m(t)$ ($m<0$):
\begin{equation}
  \label{2.6}
  \widetilde B_m(t) = B_m(t) + B_{-m}(t), \quad m > 0
\end{equation}
and for uniformity of notations we set $\widetilde B_0(t) \equiv
B_0(t)$.

Now in accordance with our assumptions, convert one equation
\eqref{2.3} in a system of cycling equations for the partial
amplitudes $a_k(t)$. Equation for $a_0(t)$ contains in the
right-hand side only function $\widetilde B_0(t)$
(see~\eqref{2.2}):
\begin{equation}
  \label{2.7}
  \dot a_0 = - i E a_0 - C^2 \int\limits_0^t
  \widetilde B_0 (t - t') \, a_0(t') \, {\rm d}t',
  \quad a_0(0) = 1.
\end{equation}
and in accordance with the above-mentioned arguments, an equation
for $a_k(t)$ ($k>0$) can be rewritten:
\begin{equation}
  \label{2.8}
  \dot a_k = - i E a_k - C^2 \int\limits_0^t
  \sum\limits_{m=0}^{k} \widetilde B_m(t - t') \,
  a_{k-m}(t') \, {\rm d} t', \quad
  a_k(0) = 0 \,\, (k > 0).
\end{equation}
It is notable that system \eqref{2.8} is accurate.

The Laplace transformation of \eqref{2.7} gives the following
expression for $a_0(p)$:
\begin{equation}
  \label{2.9}
  a_0(p) = \dfrac1{p + iE + C^2 \, \widetilde B_0(p)}.
\end{equation}
Here $\widetilde B_0(p)$  is the Laplace transform of $B_0(t)$
(see~\eqref{2.2}):
\begin{equation}
  \label{2.10}
  \widetilde B_0(p) = \dfrac12 \left( \sqrt{p^2 +4} - p \right).
\end{equation}
Recall that expression \eqref{2.9} is nothing else then amplitude
$a(t)$ in the infinite lattice when there is no reflections.

The Laplace transformation of system \eqref{2.8} gives the
algebraic system of recurrence relationships for $a_k(p)$ ($k>0$):

\begin{equation}
  \label{2.11}
  a_k(p) = -C^2 \, a_0(p) \sum\limits_{m=0}^{k-1}
  \widetilde B_{k-m}(p) \, a_m(p).
\end{equation}
The Laplace transform $\widetilde B_m(p)$ of the function
$\widetilde B_m(t)$  (see~\eqref{2.1}) is given by the integral:
\begin{equation}
  \label{2.12}
  \widetilde B_m(p) = \dfrac{2}{\pi} \int\limits_0^t  \sin^2(y)
  \exp[- 2 i m (N+1) y]
  \left[ \dfrac{1}{p + 2i \cos(y)} + \dfrac1{p - 2i \cos(y)} \right]
  {\rm d} y.
\end{equation}

\begin{figure}
\begin{center}
  \includegraphics[width=60mm]{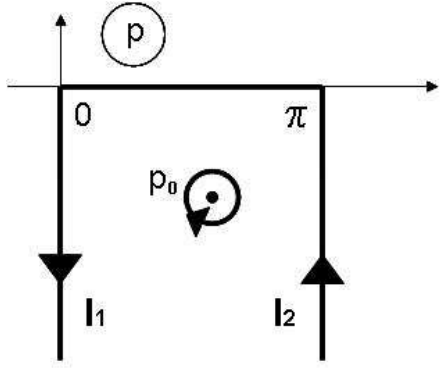}
\end{center}
 \caption{
 \label{Fig_01b}
The integration along the path $[0, \, \pi]$ in expression
\eqref{2.12} is changed to the residue in the pole $p_0$ and two
integrals $I_1$ and $I_2$.
  }
\end{figure}

An explicit form for the integral \eqref{2.12} can be obtained.
With this aim in view, the integration contour should be deformed
as shown in Fig.~\ref{Fig_01b}.
As the integrand in \eqref{2.12} has period $\pi$, integrals $I_1$
and $I_2$ are cancelled as they have opposite integration paths.
Consequently $\widetilde B_m(p)$ is defined only by the pole
contribution at the point $y_0$ ($\cos (y_0)=ip/2$). Then we have:
\begin{equation}
  \label{2.13}
  \widetilde B_m(p) = \sqrt{4 + p^2}
  \left[ i \widetilde{B}_0(p) \right]^{2m(N+1)}.
\end{equation}
Due to the fact that $\widetilde{B}_m(p)$ forms the geometrical
progression by $m$, it is possibile to make an explicit summation
in the recurrence formulaes \eqref{2.11}. It is the reason why the
total amplitude $a(t)$ can be represented in the form of the
closed expression (this is done in Appendix).

The Laplace transforms of the partial amplitudes $a_k(p)$  ($k>0$)
are expressed as
\begin{equation}
  \label{2.14}
  a_k(p) = - a_0(p) \, d \, (1 - d)^{k-1}
  \left( i \widetilde{B}_0 \right)^{2k(N+1)}, \quad {\rm where}
   \quad d \equiv a_0 \, C^2 \sqrt{p^2 + 4}.
\end{equation}

To get explicit expressions for the partial amplitudes, the
inverse Laplace transform should be made. Transforming the Laplace
integral to the integration path around the cut $[-2i, 2i]$, one
can get an expression for $a_0(t)$:
\begin{equation}
  \label{2.15}
  a_0(t) = \dfrac1{\pi} \int\limits_{-2}^2
  \exp(i x t) \, {\rm Im} \left\{
  \left[
        { x \left( 1 - \dfrac{C^2}{2} \right)  + E -
    \dfrac{i C^2}{2} \sqrt{4 - x^2} }
  \right]^{-1} \right\} {\rm d} x .
\end{equation}
Further on we get the following expression for the partial
amplitudes $a_k(t)$:
\begin{equation}
  \label{2.16}
  a_k(t) = -\dfrac{C^2}{\pi} \int\limits_{-2}^2
  \sqrt{4 - x^2} \exp(i x t) {\rm Re}
  \left[
  a_0^2 \left( 1 - C^2 a_0 \sqrt{4-x^2} \right)^{k-1}
  \left( i \widetilde{B}_0 \right)^{2k(N+1)}
  \right] {\rm d} x .
\end{equation}
Here the Laplace transform for $a_0(p)$ (see~\eqref{2.9} and
\eqref{2.13}) are performed at $p=ix$. An analysis of expression
for $a_k(t)$ shows that, as supposed, $a_k(t) \ll 1$ when
$t<(k+1)N$.

Therefor, if the partial sum $a_0 + a_1 + \ldots a_k$ is taken for
the representation of amplitude $a(t)$, then the error of such
approximation is of the same order as the smallness of
$\widetilde{B}_k(t)$ (see~\eqref{2.4}), $\sim [k+1)(N+1)]^{-3}$.

\begin{figure}
\begin{center}
  \includegraphics[width=100mm]{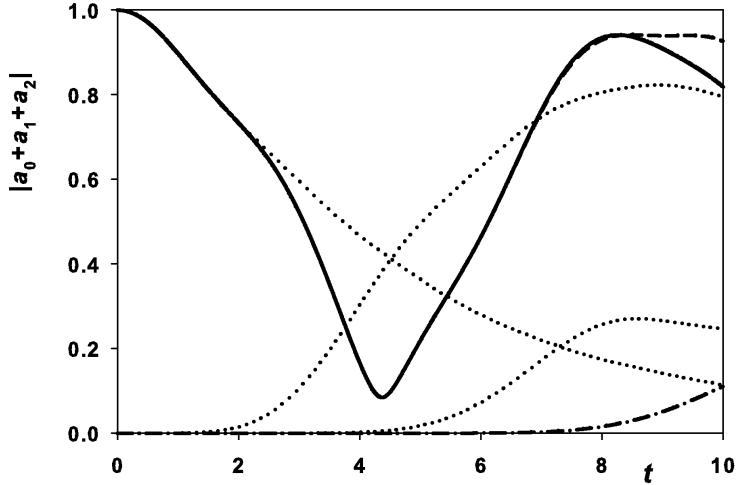}
\end{center}
 \caption{
 \label{Fig_02b}
The comparison of the limited sum of partial amplitudes $a_0 + a_1
+ a_2 $ (solid line) with the numerical integration (dashed line).
Dotted lines -- partial amplitudes $a_0$, $a_1$, $a_2$
``starting'' at times $t = 0, \, 2, \, 4$, correspondingly. Small
divergence is observe only at $t>8$ where the unaccounted partial
amplitude $a_4$ (dash-dot line) starts to make the contribution.
Parameters: $N=2, \, E=1, \, C=0.5$. Mean-square error (MSE)
$\lesssim 10^{-3}$.
         }
\end{figure}

Consider as an example very short lattice ($N=2$), and as an
approximation -- sum of only three partial amplitudes $a_0 + a_1 +
a_2 $. In Fig.~\ref{Fig_02b} this partial sum is compared with the
result of numerical integration of the Schr\"odinger equation
\eqref{1.4}. For $t<8$ the expected error is $\sim 5 \cdot
10^{-5}$. Thus the representation of $a(t)$ by the sum of partial
amplitudes is a very good approximation even for short lattices
(an accuracy increases if lattice is longer).

%


\section{Recursion. Multiple returning to the initial state.}

\begin{figure}
\begin{center}
  \includegraphics[width=100mm,angle=0]{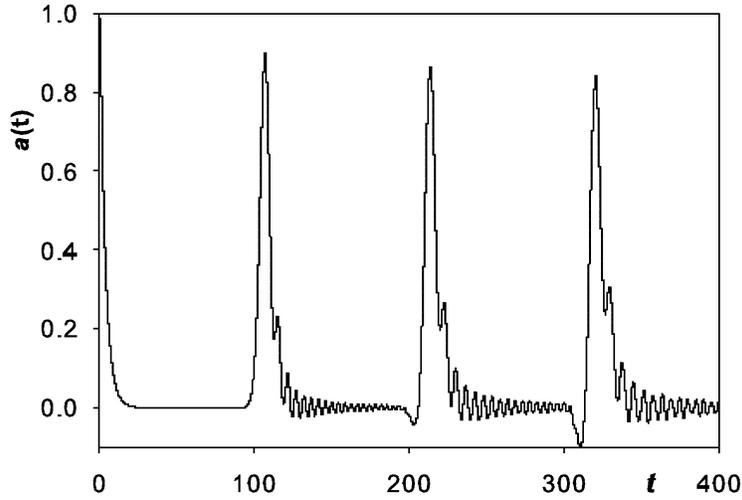}
\end{center}
 \caption{
 \label{Fig_03b}
Sum of partial amplitudes $a_0, \, a_1,\, a_2, \, a_3$ practically
coincide with the total amplitude $a(t)$. Numerical results are
not shown as they excellently coincide with analytical result (MSE
$\lesssim 10^{-4}$). Parameters: $N=100, \, E=0, \, C^2=0.25$.
  }
\end{figure}

If the lattice is long enough then partial amplitudes, following
each other, have enough time to damp on the corresponding time
range $[t \div t+N]$. In this case the partial amplitudes do not
interfere and reproduce the total amplitude with very high
accuracy (see Fig.~\ref{Fig_03b}). The maxima of returning
amplitudes slowly decrease.

\begin{figure}
\begin{center}
  \includegraphics[width=100mm,angle=0]{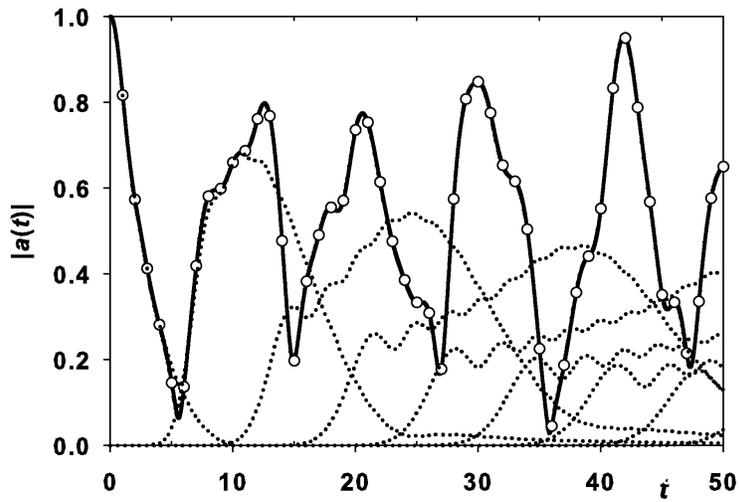}
\end{center}
 \caption{
 \label{Fig_04b}
Solid line -- sum of partial amplitudes $|a_0 + a_1 + \ldots, +
a_8|$. It practically coincides with amplitude $a(t)$. Dots --
partial amplitudes $a_0, \, a_1, \ldots, a_8$. Maximal value of
returned amplitude is $\approx 0.95$ (at $t=42$). Main
contributions to  maximum give partial amplitudes $a_2, \, a_3, \,
a_4, \, a_5, \, a_6$. Empty circles -- numerical result. MSE
$\lesssim 10^{-4}$.  Parameters: $N=10, \, E=0, \, C^2=0.4$.
  }
\end{figure}

Partial amplitudes interfere in the short lattices and maxima of
returning amplitudes are irregular. The dependence of the total
amplitude $a(t)$ vs. time for the lattice with $N=10$ is shown in
Fig.~\ref{Fig_04b}.

Numerical analysis performed at different values of parameters $C,
\, E$ и $N$ shows, that the maximal value of returned amplitude
$a_{\rm ret} \approx \! 0.972$ at $t=42$ for $N=10, \, E=0, \,
C^2=0.4$,

Incident and reflected impulses interfere on short lattices and
the degree of returning is difficult to analyze at arbitrary
parameter values $C, \, E, \, N$. But the first returning (maximal
value of the partial amplitude  $a_1$) can be treated analytically
if the lattice is long enough when amplitude $a_0$ becomes
negligible.

The expression \eqref{2.16} for $a_1(t)$ using the trigonometric
substitution of variables can be written:
\begin{equation}
  \label{3.1}
  a_1(t) = \dfrac{4 C^2}{\pi} \int\limits_{0}^{\pi}
  \sin^2(x) \, \exp[2it \cos(x)] \, {\rm Re}
  \left\{
  \dfrac{\exp[2i(N+1) x]}
  {[2 \cos(x) + E - C^2 \exp(ix)]^2}
  \right\} {\rm d} x .
\end{equation}

\begin{figure}
\begin{center}
  \includegraphics[width=100mm,angle=0]{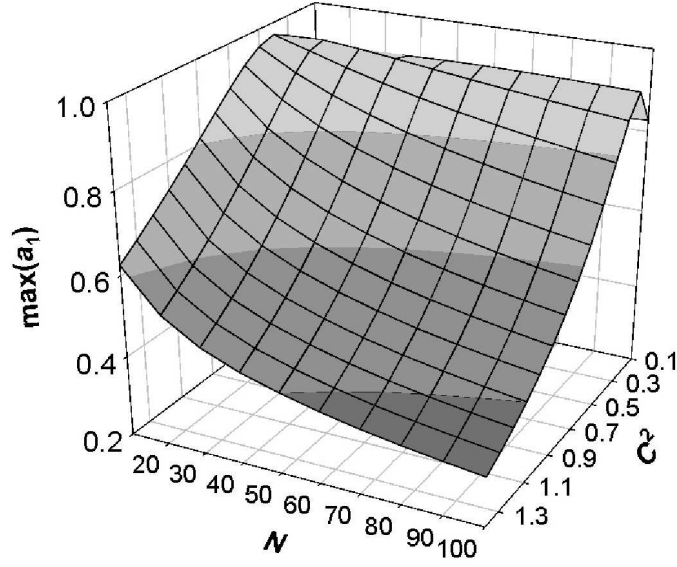}
\end{center}
 \caption{
 \label{Fig_05b}
Maximal value of amplitude of the first returning at $E=0$ and
different values of $C$ and $N$.
  }
\end{figure}

Fig.~\ref{Fig_05b} shows the maximal values of amplitude
calculated according to \eqref{3.1} at $E = 0$ and different
values of parameters $N$ and $C$. One can see that if $C^2 \approx
0.2$  then the returned amplitude practically does not depend on
the lattice length ($10<N<100$). The dissimilarity of the partial
amplitude from the total amplitude is negligible on this time
range ($N<t<2N$). Divergence becomes essential  ($\sim$15\%) for
the shortest of considered lattices ($N=10$) and smallest value of
parameter $C$ ($C^2=0.1$). Amplitude $a_0$ has no enough time to
fully decay at these parameters values.

\section{Wave packet trapping by an acceptor}

The phenomenon of multiple reflections of the wave packet
(Loschmidt echo) is unlikely to observe experimentally. The reason
is that the wave function does not interact with environment.
Below we consider the problem which mimics the experiments on the
charge transfer (CN) in DNA where the wave function is
irreversibly trapped by an acceptor. And the fraction of the wave
function trapped by an acceptor is of primary interest of this
section. This quantity can be compared with the quantum efficiency
of CT.

\begin{figure}
\begin{center}
  \includegraphics[width=100mm,angle=0]{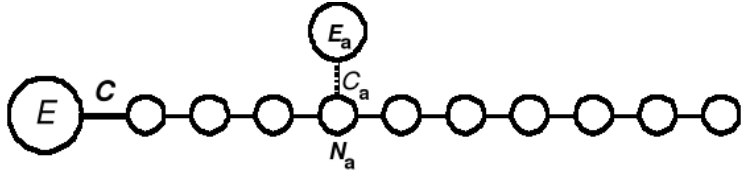}
\end{center}
 \caption{
 \label{Fig_06b}
Schematic representation of the lattice with acceptor. Acceptor is
attached to the lattice site with number $N_{\rm a}$ and has the
on-site energy $E_{\rm a}$ and hopping integral (the interaction
energy with the lattice) $C_{\rm a}$.
  }
\end{figure}

Consider the lattice with the attached site (acceptor, see
Fig.~\ref{Fig_06b}). The number of this site is $N_{\rm a}$. The
acceptor has on-site energy $E_{\rm a}$ and the hopping integral
$C_{\rm a}$. Amplitude of the wave function on the acceptor is
labelled by $b_{\rm a}$. Initially we limit ourself by the weak
bounding energy between the lattice and acceptor, i.e. $C_{\rm a}
\ll 1$. This approximation allows to make necessary analytical
estimations.

The system of equations \eqref{1.4} changes in an obvious way: an
equation for the amplitude of the wave function on the acceptor is
added:
\begin{equation}
  \label{1.a}
  i \dot b_{\rm a}(t) = E_{\rm a} b_{\rm a}(t) + C_{\rm a} b_{N_{\rm
  a}}, \quad b_{\rm a}(t=0) = 0.
\end{equation}
The equation for the site $N_{\rm a}$ is also modified:
\begin{equation}
  \label{2.a}
  i \dot b_{N_{\rm a}}(t) = b_{N_{\rm a} - 1} + b_{N_{\rm a} + 1}
  + C_{\rm a} b_{\rm a}.
\end{equation}
Other equations stay unchanged.

\begin{figure}
\begin{center}
  \includegraphics[width=100mm]{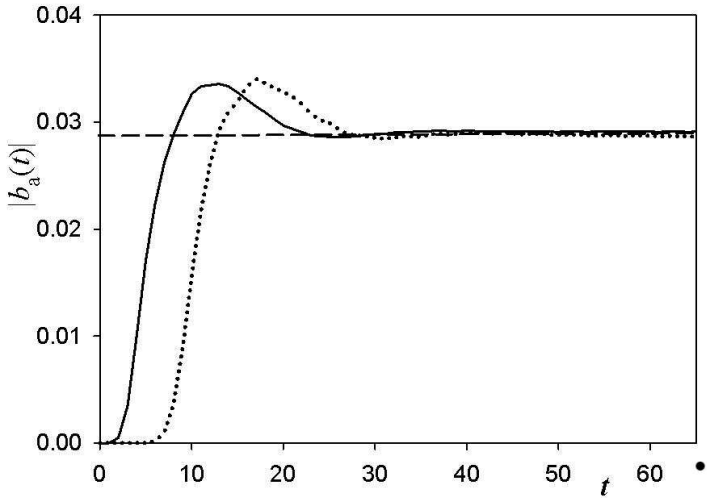}
\end{center}
 \caption{
 \label{Fig_07b}
The dependence of the wave function amplitude on the acceptor
$|b_{\rm a}(t)|$ for two positions of the acceptor on the lattice:
$N_{\rm a} = 5$ (solid line) and $N_{\rm a} = 15$ (dots). Dashed
line -- expression \eqref{7aa}. Parameters: $N = 100, \,\, C =
0.4, \,\, E = 0.5, \,\, C_{\rm a} = 0.02, \,\, E_{\rm a} = 0.3$.
  }
\end{figure}

Staying in the frameworks of the initially formulated problem,
consider now results obtained in numerical simulation in the case,
when the acceptor is located close to the impurity site (donor).
It turns our that the fraction of the wave function on the
acceptor, being captured, stays on the acceptor for a long time.
Fig.~\ref{Fig_07b} shows this phenomenon for the lattice
consisting of $N=100$ sites. (The time range is such, that the
reflected impulse has no time to return back after reflection).

It is possible to estimate the dependence of the wave function
amplitude on the acceptor vs. time in the approximation of the
weak coupling ($C_{\rm a} \ll 1$). Note that in this approximation
the acceptor affects the lattice very weakly. Therefor the lattice
is not disturbed and it is described by  Eq.~\eqref{1.4}.
Amplitude of the wave function on the acceptor will be calculated
according to the obvious expression resulting from \eqref{1.a}:
\begin{equation}
  \label{2aa}
  b_{\rm a}(t) = - i C_{\rm a} \exp(- i E_{\rm a} t)
  \int\limits_0^t \exp(i E_{\rm a} \tau) b_{N_{\rm a}}
  \, {\rm d} \tau .
\end{equation}

\begin{figure}
\begin{center}
  \includegraphics[width=100mm,angle=0]{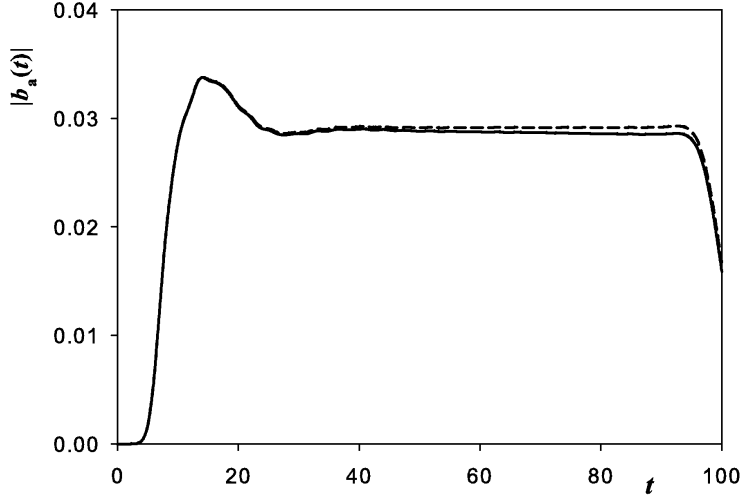}
\end{center}
 \caption{
 \label{Fig_08b}
A comparison of the accurate and approximate solutions for the
acceptor attached to the tenth lattice site ($N_{\rm a}=10$).
Solid line -- accurate solution (an influence of the acceptor on
the lattice is accounted). Dashed line -- unperturbed lattice.
Parameters: $N = 100, \,\, N_{\rm a}=10, \,\, C = 0.4, \,\, E =
0.5, \,\, C_{\rm a} = 0.02, \,\, E_{\rm a} = 0.3$.
  }
\end{figure}

Fig.~~\ref{Fig_08b} shows the comparison of two solutions of the
Schr\"odinger equation: accurate (expression \eqref{1.4} ) and
approximate (equation \eqref{1.a}). One can see that these
solutions differ very little and an approximation by the
unperturbed lattice is very good.

Analytically will be considered the case when time is large enough
such that amplitude $a(t)$ on the impurity site decreased
practically to zero. It means that the lattice is long and time is
large, and the impulse and its tail went away from the acceptor.
Then the lattice can be considered as having infinite length.

As is seen from \eqref{2aa}, it is necessary to evaluate integral
$\int_0^t \exp(i E_{\rm a} \tau) b_{N_{\rm a}} \, {\rm d} \tau$
(the phase multiplier $\exp(- i E_{\rm a} t)$ is unessential for
the modulus of the wave function). We consider the amplitude on
acceptor in the limit $t \to \infty$.

To evaluate the integrals, system \eqref{1.4} should be multiplied
by $\exp(iEt)$ and integrated in the limits from 0 to $\infty$.
Lets introduce the notations:
\begin{equation}
  \label{3a}
  I_0 \equiv \int\limits_0^{\infty} \exp(i E_{\rm a} \tau)
  a(\tau) {\rm d} \tau, \qquad
I_k \equiv \int\limits_0^{\infty} \exp(i E_{\rm a} \tau)
  b_k(\tau) {\rm d} \tau.
\end{equation}
Then for $I_k$ we get the recurrence relations:
\begin{equation}
  \label{4a}
  \begin{split}
  C I_1 = & \, -i + (E - E_{\rm a}) I_0 \\
    I_2 = & \,\, E_{\rm a} I_1 - C I_0 \\
    I_3 = & \,\, E_{\rm a} I_2 - I_1 \\
    I_4 = & \,\, E_{\rm a} I_3 - I_2 \\
  \ldots   & \,\, \ldots \, \ldots
  \end{split}
\end{equation}

Amplitude $a(t)$ on the impurity site in the considered
approximation is $a_0(t)$ and integral $I_0$ is the Laplace
transform $a_0(p) \,\, (p = iE_{\rm a})$. As the result we get:
\begin{equation}
  \label{5a}
  I_0 = \left[ i (E - E_{\rm a}) + C^2 \exp(i \phi) \right]^{-1},
  \qquad \phi = \arcsin(E_{\rm a}/2).
\end{equation}
System of equations \eqref{4a} has the following solution:
\begin{equation}
  \label{6a}
  I_k = C I_0 \left[-i \exp(i \phi) \right]^k .
\end{equation}
Thus the limiting values of the acceptor amplitudes (with the
accuracy of oscillating multiplier $\exp(-iEt)$) on different
sites are$- i C_{\rm a}  I_k$ (see \eqref{2.a}) and differ only by
phase multiplier. In this case the amplitude on acceptor is
\begin{equation}
  \label{7a}
  b_{\rm a}(t) \approx -i C_{\rm a} C \exp(-i E_{\rm a} t)
   I_0 \, \exp[-i \exp(i \phi)]^k .
\end{equation}

If the phase multipliers, unessential for the amplitude of the
wave function, are eliminated, then the amplitude is
\begin{equation}
  \label{7aa}
  b_{\rm a}(t) \approx
  \dfrac{C_{\rm a} C}{i(E - E_{\rm a}) + C^2 \exp(i \phi)}, \qquad
  \phi = \arcsin(E_{\rm a}/2).
\end{equation}

Fig.~\ref{Fig_07b} demonstrates that the limiting value of
amplitude coincides with the numerical calculation.

It follows from \eqref{7aa} that at fixed values of the hopping
integrals $C$ and $C_{\rm a}$, maximal value of $b_{\rm a}$ is
achieved in ``resonance'' values of $E$ and $E_{\rm a}$, when $E =
E_{\rm a}$. In this resonance case $b_{\rm a}(t) \approx C_{\rm
a}/C$.

\begin{figure}
\begin{center}
  \includegraphics[width=90mm]{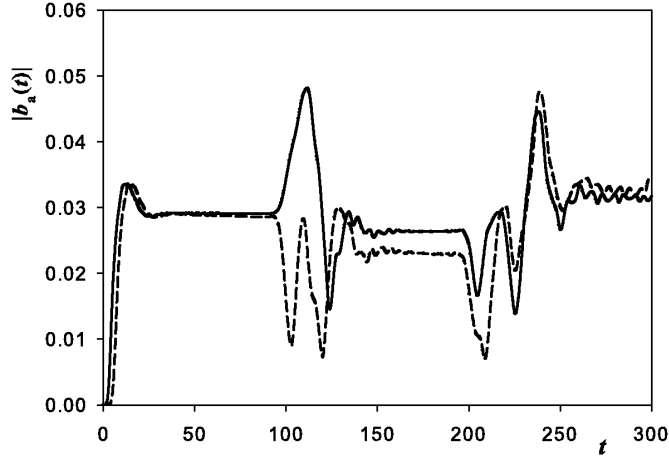}
\end{center}
 \caption{
 \label{Fig_09b}
Amplitude of the wave function on acceptor when $N_{\rm a}=5$
(solid line) and $N_{\rm a}=10$ (dashed line). Time is such that
the double reflection occurs. Parameters: $N = 100, \,\,  C = 0.4,
\,\, E = 0.5, \,\, C_{\rm a} = 0.02, \,\, E_{\rm a} = 0.3$.
  }
\end{figure}

If times are such that the impulse reflects, returns and passes by
the acceptor, then the amplitude variations are irregular and
depend on the acceptor location on the chain
(see~Fig.\ref{Fig_09b}).

There was considered above the cases, when the bounding of the
acceptor with the lattice is weak. But practically, for the
efficient charge transfer, it is necessary to obtain the
conditions when the degree of the CT is higher, i.e. parameter
$C_{\rm a}$ should be larger. The value $E_{\rm a}$ also plays
some role.

\begin{figure}
\begin{center}
  \includegraphics[width=90mm,angle=0]{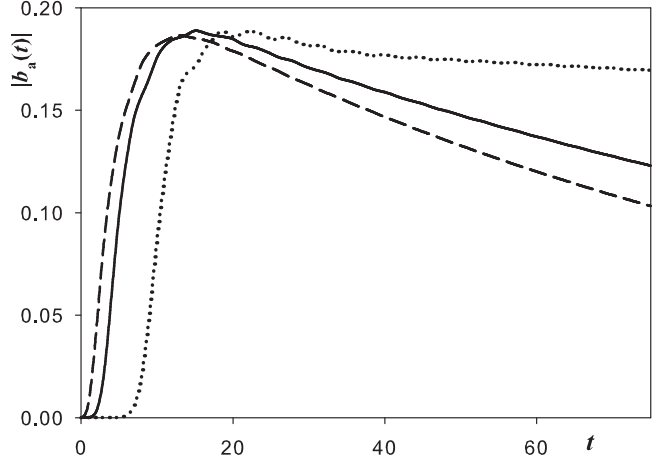}
\end{center}
 \caption{
 \label{Fig_10b}
Dependence of the wave function amplitude on the acceptor vs. time
for few locations of the acceptor on the lattice. $N_{\rm a} = 2$
(dashed line), $N_{\rm a} = 5$ (solid line) и $N_{\rm a} = 15$
(dots). Parameters: $N = 100, \,\,C = 0.5, \,\, E = 0.5, \,\,
C_{\rm a} = 0.1, \,\, E_{\rm a} = 0.5$. (According to \eqref{7aa}
the value $b_{\rm a} = 0.2$).
  }
\end{figure}

Below we consider few examples when parameter $C_{\rm a}$ has
comparatively large value. Fig.~\ref{Fig_10b} shows the dependence
of the wave function amplitude on the acceptor, when the hopping
integral $C_{\rm a} = 0.1$. The amplitude becomes well larger and
reaches value $|b_{\rm a}| \lesssim 0.2$.

\begin{figure}
\begin{center}
  \includegraphics[width=90mm,angle=0]{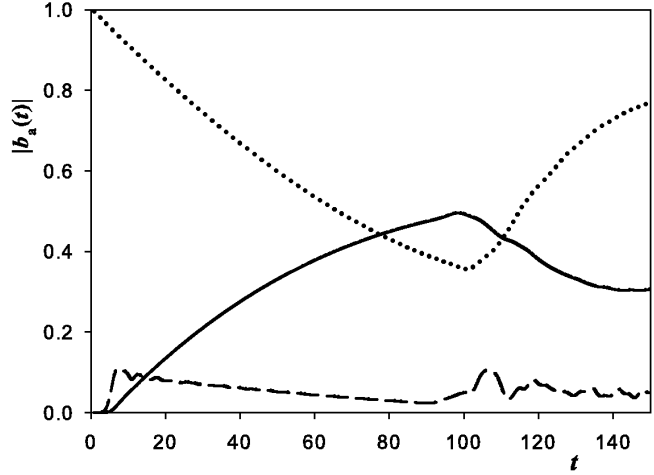}
\end{center}
 \caption{
 \label{Fig_11b}
Dependence of the wave function amplitudes vs. time  on the
acceptor $|b_{\rm a}(t)|$ (solid line), on the impurity site
$|a(t)|$ on the left lattice end (dots) and on the acceptor site
$|b_{10}(t)|$ (dashed line). Parameters: $N = 100, \,\,C = 0.1,
\,\, E = 0.5, \,\, C_{\rm a} = 0.1, \,\, E_{\rm a} = 0.5$.
  }
\end{figure}

In the case of the total resonance (when  $E = E_{\rm a}$ and $C =
C_{\rm a}$), amplitude  $|b_{\rm a}|$ becomes even more and this
case is shown in Fig.~\ref{Fig_11b}.

From two latter figures it follows that the acceptor population
can change significantly depending on the parameters. When the
population is small (Fig.~\ref{Fig_10b}) then time of life is
comparatively large. And on contrary, time of life is small when
population is large (Fig.~\ref{Fig_11b}). This peculiarity has
natural explanation: the larger is the acceptor bounding with the
lattice the shorter is time of life.

\begin{figure}
\begin{center}
  \includegraphics[width=90mm,angle=0]{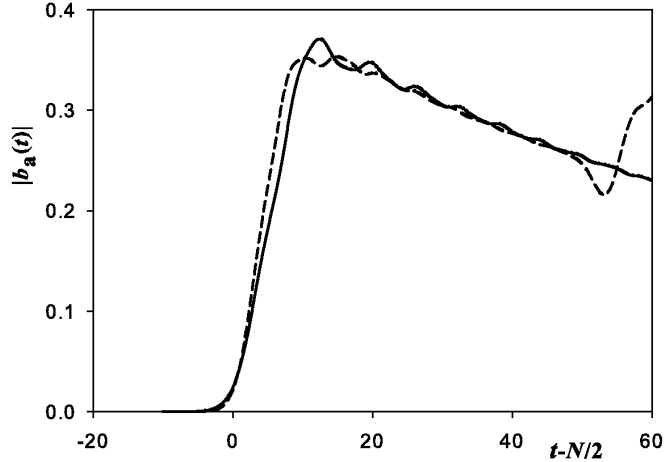}
\end{center}
 \caption{
 \label{Fig_12b}
Dependence of the wave function amplitude on the acceptor (located
on the right lattice end) vs. time for two lattice lengths: $N =
100$ (solid line) and $N = 50$ (dashed line). The time point of
reference is shifted back by the value $N/2$ for the data
comparison. Parameters: $C = 0.5, \,\, E = 0.5, \,\, C_{\rm a} =
0.1, \,\, E_{\rm a} = 0.5$.
       }
\end{figure}

And finally we consider the case when an acceptor is located on
the right lattice end ($N$th site is an acceptor) (see
Fig.~\ref{Fig_12b}). It is seen that the amplitude on acceptor is
rather large ($|b_{\rm a}| \approx 0.35$), and, what is very
important, does not depend on the lattice length. Moreover, time
dependencies for the amplitude decay are practically identical.
This result is in good agreement with experiments on the charge
transfer in synthetic DNA and polypeptides where the CT
probability does not depend on the distance.

An estimation of typical time scale is necessary. It has to be
done to understand how long the wave function stays in the bounded
state, and is this time enough for photophysical or
electrochemical response for the charge registration. The typical
dynamical time (period of one vibration) is $[t]_{\rm d} \approx
1.7 \cdot 10^{-13}$ s \cite{Con00, Con03}. The typical electronic
time (time unit in this work) is approximately two orders of
magnitude shorter $[t]_{\rm e} \approx 2.2 \cdot 10^{-15}$ s
\cite{Ast12a}. As is shown above, the wave function can stay on
the acceptor during dozens time units, what is $\sim$ps. In many
cases this time is enough for effective charge trapping by an
acceptor with following registration.


\section{Conclusions}

In two papers we thoroughly analyzed the quantum dynamics of the
excitation (electronic wave function) propagation (first part),
reflection and trapping (second part). The system consists of the
homogeneous one-dimensional lattice with the impurity site, and an
excitation initially is totally localized on the impurity site.

A rather unexpected results consists in the fact that initially
localized wave function starts to move spontaneously forming well
defined wave packet. After first reflection wave packet is again
concentrated on the impurity site with the amplitude $\gtrsim
90$\% of initial amplitude. This process repeats many times.

To describe multiple reflections of the wave packet an useful
approach consisting in the representation of the full wave
function on the impurity site $a(t)$ through the partial
amplitudes $a_k(t)$.

The temporal evolution of the wave function is described with very
high accuracy up to dozens reflection. The interference of falling
and reflected impulses occurs after these large times, which is
difficult to take into account analytically. The behavior of the
quantum dynamical system is regular in this time range. The
behavior on large times needs further detailed consideration.

Results on the wave function trapping by an acceptor can explain
recent results on the efficient ballistic charge transport in
synthetic DNA and polypeptides.


\appendix

\section{}

Original equation for the amplitude $a(t)$ has the form (see
\eqref{1.2}):
\begin{equation}
  \label{P.1}
  \dot a = - i E a - \int\limits_0^t B^N(t-\tau) \,
  a(\tau) {\rm d} \tau, \quad a(t=0) = 1.
\end{equation}
The solution of this equation for the Laplace transform $a(p)$ is:
\begin{equation}
  \label{P.2}
  a(p) = \dfrac{1}{p + i E + B^N(p)}.
\end{equation}

The Poisson representation for the kernel $B^N(p)$ has form (see
\eqref{2.1}, \eqref{2.10}, \eqref{2.13}):
\begin{equation}
  \label{P.3}
  \begin{split}
     B^N(p) = & \, C^2 \sum\limits_{m=0}^{\infty}
     \widetilde B_m(p) \\
     \widetilde B_0(p) = & \, \dfrac12
     \left(  \sqrt{p^2 + 4} - p \right) \\
     \widetilde B_m(p) = & \,
     \sqrt{p^2 + 4} \, [b(p)]^{2m(N+1)}, \quad m > 0 \\
  \end{split}
\end{equation}
where the notation
\begin{equation}
  \label{P.4}
  b(p) \equiv \left( i \widetilde B_0(p)  \right)^{2k(N+1)}
\end{equation}
is introduced. As values $\widetilde B_m$ form the geometrical
progression, we get:
\begin{equation}
  \label{P.5}
  B^N(p) = C^2 B_0(p) + C^2 \sqrt{p^2 + 4} \dfrac{b(p)}{1 - b(p)}.
\end{equation}
And the final expression for the total amplitude is:
\begin{equation}
  \label{P.6}
  a(p) = a_0 \dfrac{1-b(p)}
  {1 + b(p) \left( C^2 a_0(p) \sqrt{p^2 + 4} - 1 \right)}.
\end{equation}

An expansion into series in terms by $b(p)$ gives Laplace
transforms of partial amplitudes. By this means it was not
necessary to construct the system of the recurrence relations for
the partial amplitudes, but simply use the expansion $a(p)$ (see
\eqref{P.6}) into series in terms by $b(p)$. But this approach is
valid only for the considered model where the necessary summation
is possible and the compact expression for the kernel $B^N$ can be
written. The method suggested in the paper of the expansion into
series by partial amplitudes can be applied in other problems.

The back Laplace transformation gives the desired expression for
$a(t)$:
\begin{equation}
  \label{P.7}
  a(t) = \dfrac{1}{2 \pi} \int\limits_{-\infty}^{\infty}
  a(p) {\rm d} p, \quad p = \Delta + i \omega, \quad
  \Delta > 0.
\end{equation}
Numerically it was verified that expression \eqref{P.6} gives
accurate results.

It worth noting one intriguing property. If the integral
\eqref{P.7} is closed around the cut $[-2i,\, 2i]$, then the
result (difference of integrals taken along banks of the cut) is
zero. It was found numerically. And it follows that the function
$a(p)$ has poles in the complex plane. And amplitude $a(t)$ can be
obtained as the sum of residues in these poles. Additional
analysis is necessary to throw light on this fact.


\end{document}